\documentclass{aa}  
\usepackage{graphicx}
\usepackage{txfonts}
\usepackage{xcolor}
\usepackage{float}
\usepackage{tablefootnote}

\authorrunning{N. L. Rossignoli et al.}
\usepackage{dblfloatfix}
\usepackage{cuted}
\usepackage{hyperref}
\usepackage{xcolor}
\usepackage{color,soul}
\definecolor{Azul1}{rgb}{0,0.086,0.6}
\hypersetup{
       colorlinks=true,
       citecolor={black},
       linkcolor={Azul1},
       urlcolor={Azul1},
   }
\begin{document} 

   \title{Crater production on Titan and surface chronology}

    \author{N. L. Rossignoli
          \inst{1,2},
          R. P. Di Sisto
          \inst{1,2},
          \and
          M. G. Parisi
          \inst{1,3}
          }
   \institute{Facultad de Ciencias Astronómicas y Geofísicas, Universidad Nacional de La Plata, Paseo del Bosque S/N (1900), La Plata, Argentina \\
   \email{nrossignoli@fcaglp.unlp.edu.ar}
   \and Instituto de Astrofísica de La Plata, CCT La Plata - CONICET - UNLP, Paseo del Bosque S/N (1900), La Plata, Argentina \and Instituto Argentino de Radioastronomía, CCT La Plata -- CONICET -- CICPBA - UNLP, CC N. 5 (1894), Villa Elisa, Argentina}

   \date{Received 15 July 2021 / Accepted 5 January 2022}
 
  \abstract
   {Impact crater counts on the Saturnian satellites are a key element for estimating their surface ages and   placing constraints on their impactor population. The {\em Cassini} mission radar observations allowed crater counts to be made on the surface of Titan, revealing an unexpected scarcity of impact craters that show high levels of degradation.
   }
   {Following previous studies on impact cratering rates on the Saturnian satellites, we modeled the cratering process on Titan to constrain its surface chronology and to assess the role of centaur objects as its main impactors. 
   }
   {A theoretical model previously developed was used to calculate the crater production on Titan, considering the centaur objects as the main impactors and including two different slopes for the size-frequency distribution (SFD) of the smaller members of their source population. A simple model for the atmospheric shielding effects is considered within the cratering process and our results are then compared with other synthetic crater distributions and updated observational crater counts. This comparison is then used to compute Titan's crater retention age for each crater diameter. 
   }
   {The cumulative crater distribution produced by the SFD with a differential index of $s_2 = 3.5$ is found to consistently predict large craters ($D > 50 $ km) on the surface of Titan, while it overestimates the number of smaller craters. As both the modeled and observed distributions flatten for craters with $D \lesssim 25 $ km due to atmospheric shielding, the difference between the theoretical results and the observational crater counts can be considered as a proxy for the scale to which erosion processes have acted on the surface of Titan throughout the Solar System age. Our results for the surface chronology of Titan indicate that craters with $D > 50 $  km can prevail over the Solar System age, whereas smaller craters may be completely obliterated due to erosion processes acting globally.
   }
   {}

   \maketitle
%

\section{Introduction}
   \label{intro}
Titan is the largest Saturnian satellite and the only satellite in the Solar System known to possess a dense atmosphere. It was discovered in 1655 by Christiaan Huygens, but most of its surface features remained veiled until 350 years later, when the {\em Cassini-Huygens} mission began its exploration of the Saturn system. Between 2004 and 2017 {\em Cassini} performed 127 close encounters with Titan, collecting data that revealed a complex world with liquid lakes, seas, and an active hydrologic cycle based on methane. In addition, in 2005 the {\em Huygens} probe completed the first landing on a satellite other than our Moon, providing in situ measurements, such as a detailed profile of the atmosphere of Titan.
Before {\em Cassini} the crater size distribution of Titan was unknown but estimated to be similar to those of the other Saturnian satellites \citep{Lopes2019}. Instead, the mission's observations uncovered an unexpectedly low number of eroded craters, indicating that geologic processes modify its surface \citep[e.g.,][]{Hedgepeth2020}. 
Following our previous studies on the mid-sized and small Saturnian satellites \citep{DiSisto2011,DiSisto2013,Rossignoli2019}, in this work we present a study of the crater production on Titan generated by centaur objects. In Sect. \ref{SecTitan}, we describe the surface of Titan and the current knowledge of its crater population. In Sect. \ref{SecMethod}, we present the method used to predict the crater distribution on Titan and its surface chronology. In Sect. \ref{SecResults}, we present our results based on the comparison with the updated crater counts and the surface age of Titan for each crater diameter. In Sect. \ref{SecConclusions}, we present our conclusions. 

\section{The surface of Titan} \label{SecTitan}
\subsection{Geological units}
Before the {\em Cassini-Huygens} mission, the composition and features of the surface of Titan were mostly unknown  \citep{Lopes2019}, although the presence of lakes or seas of liquid hydrocarbons had already been proposed based on radar observations from Arecibo \citep{Lorenz2005}. Thus, the observations made by {\em Cassini} over more than 13 years provided the first up-close and in-depth study of the satellite. The main instruments to observe the surface of Titan were the Radio Detection and Ranging (RADAR) instrument \citep{Elachi2004}, whose primary goal was to pierce through Titan's atmosphere and reveal its surface; the Visual and Infrared Mapping Spectrometer (VIMS) \citep{Brown2004}; and the Imaging Science System (ISS) \citep{Porco2004}. Data collected by these instruments helped build a global topographic map of Titan and constrain its surface properties and composition. Titan was revealed to have one of the most diverse and dynamic surfaces of the Solar System, largely altered by erosional and depositional processes that show a latitude variation \citep{Lopes2020,Hedgepeth2020}. In addition, six major geological units were identified \citep{Lopes2020}: plains, which cover 65\% of the global area and dominate the mid-latitudes; dunes, which represent 17\% of the total surface and dominate the equatorial regions ($\pm$30 latitude); hummocky terrains composed of mountain chains and isolated peaks that comprise 14\% of the global area; lakes (dry or liquid-filled), which  represent 1.5\% of Titan's total surface area and are located near the poles; labyrinths, which  have morphologies similar to karstic terrain and cover 1.5\% of Titan's total surface area; and craters, which occupy only 0.4\% of Titan's global area and are almost completely absent near the poles. Based on the location and superposition between these units, \citet{Lopes2019} were able to determine their relative ages. They conclude that the oldest units are the hummocky terrains, while the dunes  and lakes 
 are the youngest. Regarding Titan's surface composition, hummocky and crater units show lower emissivity in radiometric data consistent with water-ice materials, while plains, dunes, lakes, and labyrinths show high emissivity to radar, indicating the presence of organic materials \citep{Lopes2020}.

\subsection{Cratering counts on Titan}
\begin{figure}[t]
         {\includegraphics[width=\linewidth]{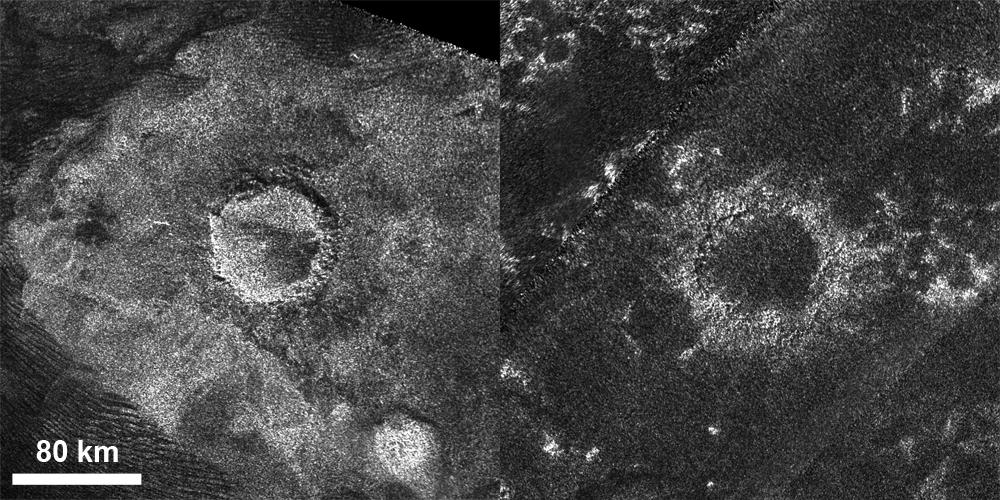}}
        \caption{{\em Cassini} RADAR images of craters Sinlap ({\em left}) and Soi ({\em right}). Image credit: PIA16638, NASA/JPL-Caltech/ASI/GSFC.} 
        \label{degra}
\end{figure}

At the end of the {\em Cassini} mission,  $\sim69\% $ of the surface of Titan was mapped by synthetic aperture
radar (SAR) \citep{Hedgepeth2020}. In this operating mode, radar images with spatial resolutions as low as 350 m were obtained \citep{Lopes2019}. Overall, the surface of Titan was revealed to present a scarcity of impact craters that is consistent with a heavily eroded surface \citep{Neish2012}. In addition, many of the identified craters show evidence for extensive modification by erosive processes, such as fluvial erosion and aeolian infill \citep[e.g.,][]{Neish2013,Lopes2019}. For example, craters Sinlap and Soi are both $\sim 80$ km in diameter, but show different degradation states (Fig. \ref{degra}). Sinlap appears to be young while Soi is extremely degraded.  \citet{Neish2013} and \citet{Hedgepeth2020} studied the crater topography on Titan and compared it to that of  Ganymede, a similarly sized, airless satellite with relatively pristine craters. Their results show that Titan's crater depths are 51\% shallower than the craters on Ganymede, which is consistent with aeolian infill as the main process of crater modification. Depending on the infilling rate, \citet{Neish2013} state that many craters on Titan may remain undetected due to rapid aeolian infilling.

\begin{figure*}[h!]
 {\includegraphics[width=\linewidth]{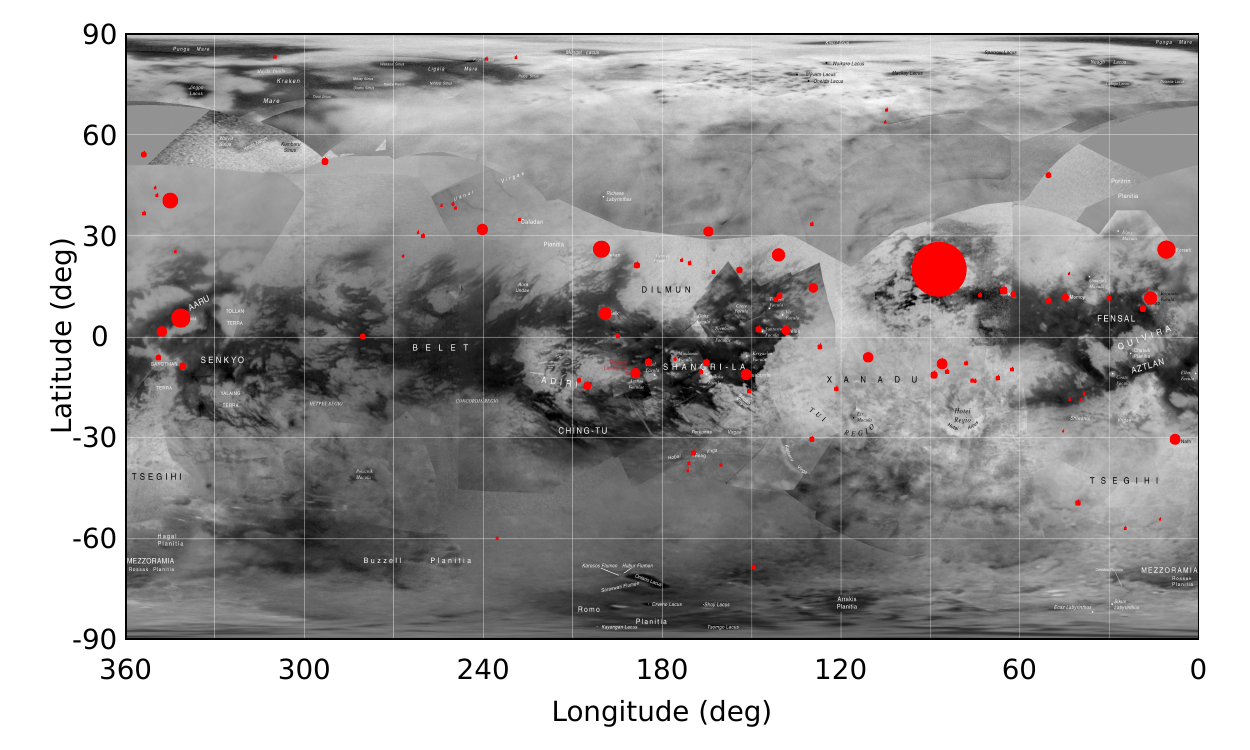}}
        \caption{Impact crater distribution over the surface of Titan from \citet{Hedgepeth2020} plotted over a global map of Titan from ISS images. Each red circle represents a crater, and its size is scaled relative to the crater diameter. Base image credit: PIA20713, NASA/JPL-Caltech/Space Science Institute/USGS.} 
        \label{mdistro}
\end{figure*}

\citet{Hedgepeth2020} reassessed the crater population obtained in previous studies \citep[e.g.,][]{Wood2010,Neish2012} using the entire SAR data set, which allowed them to identify 30 additional craters. In total, only 90 certain to possible impact craters have been identified on Titan (Fig. \ref{mdistro}), and they are not homogeneously distributed over the surface \citep{Hedgepeth2020}. In fact, 65\% of them are found within $30^{\circ}$ of the equator in a region dominated by dunes \citep{Neish2013}. 
The polar regions show a relative scarcity of craters, possibly due to the concentration of liquid lakes near the poles or the enhanced fluvial activity in the higher latitudes, which could erode craters beyond recognition \citep{Neish2016,Hedgepeth2020}.

\section{Method}
\label{SecMethod}   
In order to calculate the theoretical production of craters by centaur objects on Titan, we apply a modified version of a method previously developed in a series of works \citep{DiSisto2011,DiSisto2016,Rossignoli2019}. In addition, for this work we consider the results from our latest numerical simulation of the dynamical evolution of transneptunian objects that become centaurs \citep{DiSisto2020}. 
In this section we describe briefly the theoretical cratering method, together with the changes introduced by the new simulation and the considerations of the atmospheric effects on the impactors.

\subsection{The impactor population}

Following previous works on impact cratering models on the mid-sized and small Saturnian satellites \citep{DiSisto2013,Rossignoli2019}, we consider the main impactors to be the centaur objects. These heliocentric bodies have their origin in the transneptunian region and exhibit a transient nature due to perturbations exerted on their orbits by the giant planets.
It has been shown that the scattered disk (SD) in the transneptunian region is the subpopulation with the highest probability of having encounters with Neptune and therefore of evolving toward the planetary region of the Solar System \citep{Levison1997, DiSisto2007}.
The scattered disk objects (SDOs) that enter the giant planetary zone become centaurs and are likely to collide and produce craters on planets or their satellites. In addition, those centaur objects that enter the zone interior to Jupiter's orbit become  Jupiter-family comets (JFCs). In that region physical effects such as sublimation and splitting have important implications on the evolution of JFCs. \citet{DiSisto2009} showed that the mean physical lifetime of the JFCs is very short, on the order of a few thousand years, and that those that do survive disintegration and reenter the centaur zone, pass through it very quickly. Thus, impacts by JFCs on the planets and their satellites can be considered negligible in relation to impacts by centaurs from the SD.

In previous papers \citep{DiSisto2011, DiSisto2013, DiSisto2016, Rossignoli2019} the impact cratering rates on the mid-sized and small Saturnian satellites were modeled based on a numerical simulation of the dynamical evolution of SDOs by \cite{DiSisto2007}. In that work the authors built an intrinsic model of the SD based on the available observations at that time, and studied the contribution of the SDOs to the centaur population, which represents the main impactor population in our model. 
Since 2007, many more SDOs were discovered, which motivated a revision and an update of the model. 
Thus, \citet{DiSisto2020} built the intrinsic SD model again including new observations, which were six times as large as the 2007 sample, and analyzed the general dynamical evolution to the centaur zone, obtaining similar results as in \cite{DiSisto2007}.

\cite{DiSisto2020} also considered new estimations of the 
number and size distribution of SDOs, which have been 
improved by the recent discoveries by the Outer Solar System
Origins Survey (OSSOS) \citep{Bannister18}. It has been 
argued that the size-frequency distribution (SFD) of transneptunian objects and in particular of SDOs is not accurately modeled by a single power law, but instead presents a break at a diameter $d 
\sim$ 60--100 km, varying from a steep slope at greater
diameters to a shallow slope for smaller objects \citep[e.g.,][]{Bernstein2004, Fraser2009}. Based on the analysis of the observations of small SDOs and centaurs by OSSOS, \cite{Lawler2018} found that a break in the SFD is required
at $d \sim 100$ km (from larger to smaller SDOs). They found
a faint end slope of the absolute magnitude $H$ size
distribution of $\alpha_H$ = 0.4--0.5, which corresponds to a
differential size distribution index of $s =$ 3--3.5. At the bright end,
\cite{Elliot2005} obtained a differential size distribution
index for SDOs of $s = 4.7$. On the other hand, 
\cite{Parker2010a,Parker2010b} determined the maximum total number of SDOs with diameters larger than 100 km to be $N(d > 100 $ km$) = 3.5 \times  10^5$. Therefore, scaling the SDO population to this number and considering the SFD of SDOs with a break at $d = 100$ km and the size indexes mentioned above, we model the cumulative size distribution (CSD) of the impactor population:
\begin{xalignat}{4}
        N(>d) &= C_0 \,\bigg(\frac{1 \text{ km}}{d}\bigg)^{s_2 - 1} &&\text{for} && d \leq 100~\text{km},  \nonumber \\
        N(>d) &= \text{3.5} \times 10^{5} \, \bigg(\frac{100 \text{ km}}{d}\bigg)^{s_1-1} &&\text{for} && d > 100~\text{km}.
        \label{nr}
\end{xalignat} 
Here $C_0 = \text{3.5} \times 10^{5} \times 100^{s_2-1}$ by continuity at $d$ $=$ 100 km, $s_1 = 4.7$, and $s_2$ is modeled with two limiting values: $s_2 = 3$ and $ 3.5$, due to the uncertainty on the faint end slope of the $H$ distribution. 

An additional break to a shallower impactor SFD at diameters of $d\sim1-2$ km has been modeled in recent works, based on the cratering records on Pluto and Charon \citep[e.g.,][]{Robbins2017,Singer19}. In addition, \citet{Morbidelli2021} modeled a crater production function for Pluto, Charon, Nix, and Arrokoth based on their crater records and proposed a cumulative power law slope for the Kuiper belt objects SFD given by $N(>d) \propto d^{q_\text{KBO}} $ of $ -1.2 <  q_\text{KBO}  < -1 $ in the $0.03 \lesssim d \lesssim 2$ km range. However, in the Saturnian satellites the possibility of a break in the impactor SFD at $d\sim$1--2 km is not clear yet, given that there may be other impactor populations at play such as planetocentric objects that prevent a direct association between the observed crater distributions and the impactor population. For this reason, we have not considered this possibility in our model.

\subsection{The impact process} 
\label{crater-calc}

In order to model the impact crater size-frequency distribution on the surface of Titan produced by centaur objects, we follow the method described in our previous papers \citep[e.g.,][]{DiSisto2011, Rossignoli2019}. The number of collisions on the satellite is computed from the results of the simulation presented in \citet{DiSisto2020}, which provides the number of encounters of SDOs with Saturn. In order to relate the number of encounters within the Hill’s
sphere of the planet to the number of collisions on the
satellite, we consider a particle-in-a-box approximation 
that leads to the  equation

\begin{equation}
N_c(> d) =  \frac{v_i\,R_G^{2}}{v\,(R_{\text{H}})\,R_{\text{H}}^{2}} 8.5\,N(> d),
\label{Nc10}
\end{equation}
where $v_i$ is the relative collision velocity on Titan, v($R_{\text{H}}$) is the centaurs' mean relative encounter velocity when they enter Saturn's Hill sphere (of radius $R_{\text{H}}$), and $R_G$ is the satellite's collision radius given by $R_G=R_T\, (1+(v_{esc}/v(R_{\text{H}}))^2)^{1/2}$ to account for the gravitational focusing effect. The factor $8.5$ represents the number of encounters with Titan relative to the initial number of particles in the numerical simulation.
This number is somewhat smaller than in our previous studies \citep[e.g.,][]{Rossignoli2019}, but statistically more representative due to the larger sample of observed objects used to build the SD model.
The relative encounter velocity v($R_{\text{H}}$) is obtained from the simulation encounter files \citep{DiSisto2020}. For an airless Titan and assuming isotropic impacts, $v_i = \sqrt{v_T^2+v_0^2}$, where $v_T$ is Titan's orbital velocity and $v_0$ is the centaurs mean relative velocity when they cross the orbit of Titan. The values of these velocities are listed in Table \ref{tabladatos}, together with physical data of Titan. However, it should be noted that the atmosphere of Titan reduces both the relative velocity and the diameter of all impactors as they approach the surface. Thus, in the next section we describe how we include these effects in our cratering model.

\begin{table}
\begin{minipage}[t]{\columnwidth}
\caption{Physical parameters of Titan (mean radius $R_T$ from \citet{Zebker2009}, mass $m_T$ from \citet{Jacobson2006}, and surface gravity $g_T$) and velocities involved in the model (see Sect. \ref{crater-calc}).}
\label{tabladatos}
\centering
\renewcommand{\footnoterule}{}  
\begin{tabular}{l|c}
\hline 
\rule{0pt}{2ex}$R_T $ (km)             & 2574.73        \\
$m_T$ (g)                &     $1.345 \times 10^{26}$\\
$g_T$  (m s$^{-2}$)      &    1.35          \\  
$v_T$ (km s$^{-1}$)                 &  5.58     \\
v($R_{\text{H}}$) (km s$^{-1}$)          &     4.10 \\
$v_0$ (km s$^{-1}$)          & 8.91      \\
$v_i$ (km s$^{-1}$)     &   10.51    \\
\hline
\end{tabular}
\end{minipage}
\end{table}

\subsection{Atmospheric effects}
\label{atm}
Before the {\em Cassini-Huygens} mission, the general composition of Titan's atmosphere was  known but poorly constrained. The data collected by the Huygens Atmospheric Structure Instrument (HASI) allowed  the determination of the temperature and density profiles from an altitude of 1,400 km down to the surface \citep{Fulchignoni2005}. From this data, it was determined that the atmospheric density at the surface of Titan is approximately four times that of the Earth \citep{Neish2012}. \citet{Melosh1989} studied the minimum diameter projectile that can penetrate Titan's atmosphere at vertical incidence and computed a value for an ice impactor of  $d\sim 1.7$ km. \citet{Lorenz1997} studied the likely characteristics of the impact crater distribution on Titan before {\em Cassini} and predicted that only impactors with $d>120$ m would be able to penetrate Titan's atmosphere with a significant portion of its incident velocity. \citet{Artemieva2003} found that the atmosphere  shielded the surface from impactors smaller than 1 km and would even decelerate larger objects. Thus, in order to constrain the amount of atmospheric shielding against impactors, we consider a simple model where the effects of fragmentation, pancaking, deceleration, and ablation are included.

As the impactor traverses the atmosphere, the differential atmospheric pressure exerted on the object may lead to its fragmentation once its characteristic strength is exceeded \citep{Chyba1993}.  In this case the fragments of the disrupted object expand away from each other (the impactor ``pancakes''), but may continue to be treated as a single collective bow shock until the dissociating impactor has expanded to twice its initial radius \citep{Chyba1993,HillsGoda1993, Engel1995}, a point at which the fragments separate into individual bow shocks.   

We model the deceleration of the impactor via the conventional drag equation \citep{Engel1995}

\begin{equation}
m\,\dot{v} = -\frac{1}{2}C_D \,A\,\rho(z)\,v^2 + g_T\,m\,\sin{\alpha} , 
\label{gdrag} 
\end{equation}

\noindent where $m$, $v$, and $A=\pi d^2/4$ are the impactor's mass, relative velocity, and cross section, respectively; $g_T$ is Titan's surface gravity (see Table \ref{tabladatos}); $\rho(z)$ is the atmospheric gas density \citep{Fulchignoni2005}; and $C_D=0.64$ is the non-dimensional drag coefficient presented in \citet{Korycansky2005} (hereafter KZ05).
We consider the most probable impact angle to be $\alpha=45^{\circ}$ with respect to the horizon.
For the ablation effect, which causes continuous shedding of the impactor mass as it traverses the atmosphere, the mass variation is given by (KZ05)
\begin{equation}
\dot{m} = - C_A\,\rho(z)\,A\,v , 
\label{abl} 
\end{equation}

\noindent where $C_A=0.71$ is the non-dimensional ablation coefficient for Titan (KZ05) and the impactors are modeled following KZ05 as cylinders of constant density $\rho_{\text{i}}$ and length $h=4m/(\pi\rho_{\text{i}}d^2)$. In the present work we consider the impactors to be made of ice, thus $\rho_{\text{i}}=1$ gr cm$^{-3}$.

In order to include the pancaking effect in the model, the $W$ term is introduced, resulting in the  expression for the variation of the impactor diameter (KZ05)

\begin{equation}
\dot{d} = \frac{2\,\dot{m}}{\pi\,\rho_{\text{i}}\,d^2} + W , 
\label{dim} 
\end{equation}
 and
\begin{equation}
\dot{W} = C_p\,\frac{\rho(z)\,v^2}{\rho_{\text{i}}\,d} , 
\label{pan} 
\end{equation}
where $C_p=0.75$ is the pancaking coefficient as modeled in KZ05. Based on the models presented in \citet{HillsGoda1993}, \citet{Chyba1993}, and \citet{Artemieva2003} we follow the motion of the impactors through Titan's atmosphere considering that its effects are negligible for altitudes higher than $\sim$ 200 km from the surface. As the impactors traverse through the atmosphere they 
suffer ablation and deceleration. Disruption occurs when the impactor's strength $S=1\times10^7$ dyn cm$^{-2}$ is exceeded by the atmospheric pressure $P=\rho(z)\,v^2$ \citep{HillsGoda1993}. The resulting fragments spread while they continue their journey through the atmosphere until they reach the surface or the expanded impactor grows to twice its initial radius. In the latter case we consider a simple approach where the expanded impactor separates into two fragments and the mass of the initial impactor is divided randomly between the two fragments. The transverse dispersion velocity imparted is given by $v_{\text{t}}=(0.41\,\rho(z)/\rho_{\text{i}})^{1/2}\,v$ (KZ05) and each fragment can experience successive fragmentations.

\subsection{Crater scaling law}
\label{cr1}
In order to obtain the number of collisions on Titan as a function of the impactor diameter, we substitute Eq. (\ref{nr}) into Eq. (\ref{Nc10}). Considering that impactors are decelerated and ablated and may even be fragmented as they traverse the atmosphere of Titan (see Sect. \ref{atm}), they will reach the satellite's surface with a reduced size and velocity. Thus, the following step is to relate the final impactor diameter $d_\text{f}$ to the crater diameter it produces. 
This relation is modeled by the scaling law from \cite{Holsapple2007}, which gives the transient diameter $D_{\text{t}}$ of a crater generated by an impactor of diameter $d_\text{f}$ through the equation
\begin{equation}
D_{\text{t}} = K_{1}\left[\left(\frac{g_T\,d_\text{f}}{2\,v^{2}_{\text{f}}}\right)\left(\frac{\rho_{\text{t}}}{\rho_{\text{i}}}\right)^{\frac{2\,\nu}{\mu}}
+ K_{2}\left(\frac{Y}{\rho_{\text{t}}\,v^{2}_{\text{f}}}\right)^{\frac{2+\mu}{2}}\left(\frac{\rho_{\text{t}}}{\rho_{\text{i}}}\right)^{\frac{\nu\,(2+\mu)}{\mu}}\right]^{-\frac{\mu}{2+\mu}} d_\text{f},
\label{HH07}
\end{equation}
where $\rho_{\text{i}}=1$ g cm$^{-3}$ is the impactor density and $d_\text{f}$ and $v_{\text{f}}$ are respectively its final diameter and vertical velocity  when it reaches the surface of Titan. The parameters $\rho_{\text{t}}$, $\mu$, $\nu$, $Y$, $K_1$, and $K_2$ depend on the target material and $g_T$ is Titan's surface gravity (see Table \ref{tabladatos}). Since the outermost layer ($\sim$ 100 km thick) of the 
surface of Titan is mostly composed of cold water ice \citep{Neish2013}, we consider $\rho_{\text{t}}= 1$ g cm$^{-3}$ and continue to use the values for the icy Saturnian satellites of $\mu=0.38$, 
$\nu=0.397$, and $K_1=1.67$, while we select the value of $K_2=0.8$. For the target strength $Y$ we adopt the most commonly used value of the tensile strength of polycrystalline water ice: $Y=1\times10^7$ dyn\,cm$^{-2}$ \citep{Manga2007}.

With Eq. (\ref{HH07}) one can determine from the dominant term if, during the impact crater formation, the crater growth is limited by the target's gravity (first term) or its strength (second term). In the case of Titan, all craters are formed under the gravity regime. The initial compression and excavation stages of the impact process define the transient crater diameter $D_{\text{t}}$ given by Eq. (\ref{HH07}). Then, a final stage of gravity-driven crater collapse takes place and expands the crater to its final diameter \citep[e.g.,][]{Collins2012}. This last stage will affect crater sizes differently depending on the initial impact energy, which separates craters into two morphological categories. 
Smaller and bowl-shaped craters are called simple craters and tend to present a depth to diameter ratio near 1:5 \citep{Melosh1999}. In contrast, above a certain crater diameter a complex crater is formed, with a central peak or ring, a relatively flat floor, and a smaller depth-to-diameter ratio. The transition from a simple to a complex morphology occurs at a distinct crater size given by \citep{Kraus2011} 
\begin{equation}
D^{*} = \frac{g_{gan}}{g_T}\,2\,R^{*}_{gan}, 
\end{equation}
where  $g_{gan}=1.43$ m\,s$^{-2}$ is the surface gravity of Ganymede and $R^{*}_{gan}=1 $ km is the radius of the transition crater between simple and complex craters for Ganymede \citep{Schenk2002}. Thus, following the relation presented in \citet{Kraus2011}, the final crater diameter can be obtained with

\begin{xalignat}{4}
 D &= (1.3\, k)D_t  &&\text{for} && D_t \leq D^*/1.3\,k,  \nonumber \\
 D &= D_t\,(1.3\,k)^{1/(1-\eta)} \left(\frac{D_t}{D^{*}}\right)^{\eta/(1-\eta)}   &&\text{for} &&  D_t > D^*/1.3\,k,
\label{dcrater}
\end{xalignat} 
where $k=1.19$ and $\eta=0.04$ \citep{Kraus2011}.

\subsection{Surface age}  
\label{surf-age}
\begin{figure}[t]
\includegraphics[width=1\linewidth]{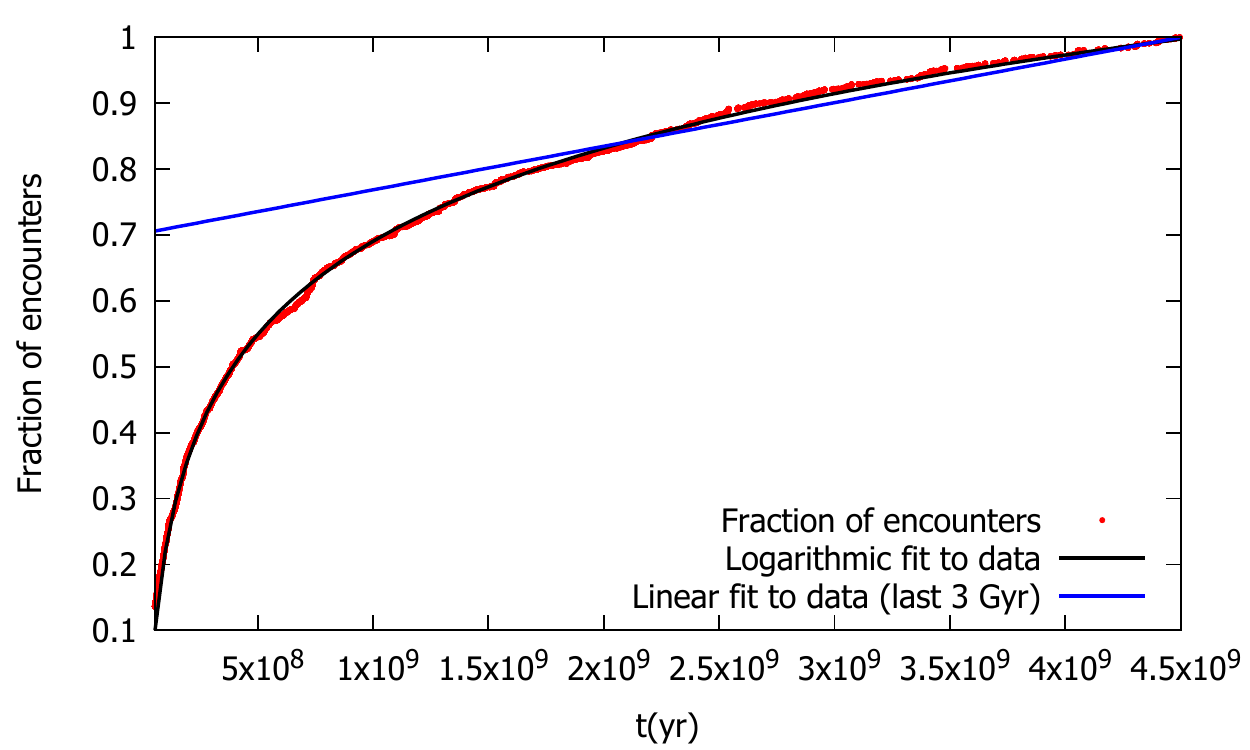}
\caption{Cumulative fraction of encounters of centaurs with Saturn as a function of time (red dots), together with a logarithmic fit to data (black solid line) and a linear fit to data for the last 3 Gyr (blue solid line).}
\label{encounters}
\end{figure}

Craters on the surface of Titan are scarce and present different degradation states (Fig. \ref{degra}). 
Considering that erosive processes act at a global scale on the satellite and may even be able to modify craters beyond detection \citep{Neish2016}, we study Titan's surface chronology. As in previous works \citep{DiSisto2016,Rossignoli2019}, we adopt a simple approach to constrain the global effect of erosive processes based on the difference between our simulated cratering counts and the observed ones. Thus, the surface chronology calculated in this work represents Titan's crater retention age, which is a measure of the extent to which erosional processes have been able to erode craters beyond detection limits in {\em Cassini} radar data.
Following the method developed in \cite{DiSisto2016}, we obtain the cratering time dependence considering Eq. (\ref{Nc10}), where the cumulative number of craters on Titan is proportional to the number of encounters of centaurs with Saturn. Thus, the dependence of cratering with time is the same as that of the encounters of centaurs with Saturn (see \citealt{DiSisto2011}). Based on the results of the simulation from \citet{DiSisto2020}, Fig. \ref{encounters} shows the cumulative number of encounters of centaurs with Saturn for a given time, normalized to the total number of encounters of centaurs with Saturn over the entire simulation.  
As  can be seen, the simulation data follows a logarithmic behavior, and can be fitted by the function
$F(t) = a\,ln(t)+b$, where $a=0.20412$ $\pm$ 0.00002, $b=-3.5398$ $\pm$ 0.0005, and $t$ is in years. In addition, Fig. \ref{encounters} shows a linear fit to data for the last 3 Gyr of the simulation, given by $ G(t) = \dot{C}(t-4.5\times10^9)+1$, where $\dot{C}=6.61567\times10^{-11}$ $\pm 7.6\times10^{-14}$ and $t$ is in years.

In order to obtain the theoretical cumulative number of craters for a given time, we use the logarithmic fit:
\begin{equation}
N_c(> D,t)=F(t) \, N_c(> D) 
\label{time} 
.\end{equation}
In Eq. (\ref{time}), if $t$ equals the age of the Solar System, we can obtain the expected cumulative number of craters on Titan over the age of the Solar System for the case where the satellite was not affected by erosive processes strong enough to be capable of erasing crater evidence beyond detection. Comparing these results with the observed crater counts allows us to calculate Titan's surface age $\tau(>D)$ as a function of each crater diameter, which is the maximum crater retention age for each crater diameter according to our model:

\begin{equation}
\tau(>D) = t_f\,\left(1-e^{-\frac{N_o(>D)}{a\,N_c(>D)}} \right) . 
\label{agesup} 
\end{equation}
Here $t_f$ = 4.5 Gyr is the age of the Solar System, $N_0(>D)$ is the satellite's cumulative number of observed craters for each crater diameter, and $N_c(>D)$ is our model's cumulative number of craters for each crater diameter.

\section{Results}
\label{SecResults}   
In the previous section we  describe the method used to calculate the crater distribution on Titan produced by centaur objects over the age of the Solar System.
Given the uncertainty in the size distribution for the smaller objects of the centaur source population, in this section we present our results for two limiting values of the $s_2$ index in Eq. (\ref{nr}), $s_2=3$ and $s_2=3.5$. Our predicted crater distributions are compared with the updated observational crater counts presented in \citet{Hedgepeth2020}, where the authors reassessed Titan's crater population using the entire {\em Cassini} SAR data set.     

\begin{figure}[h!]
\includegraphics[width=1\linewidth]{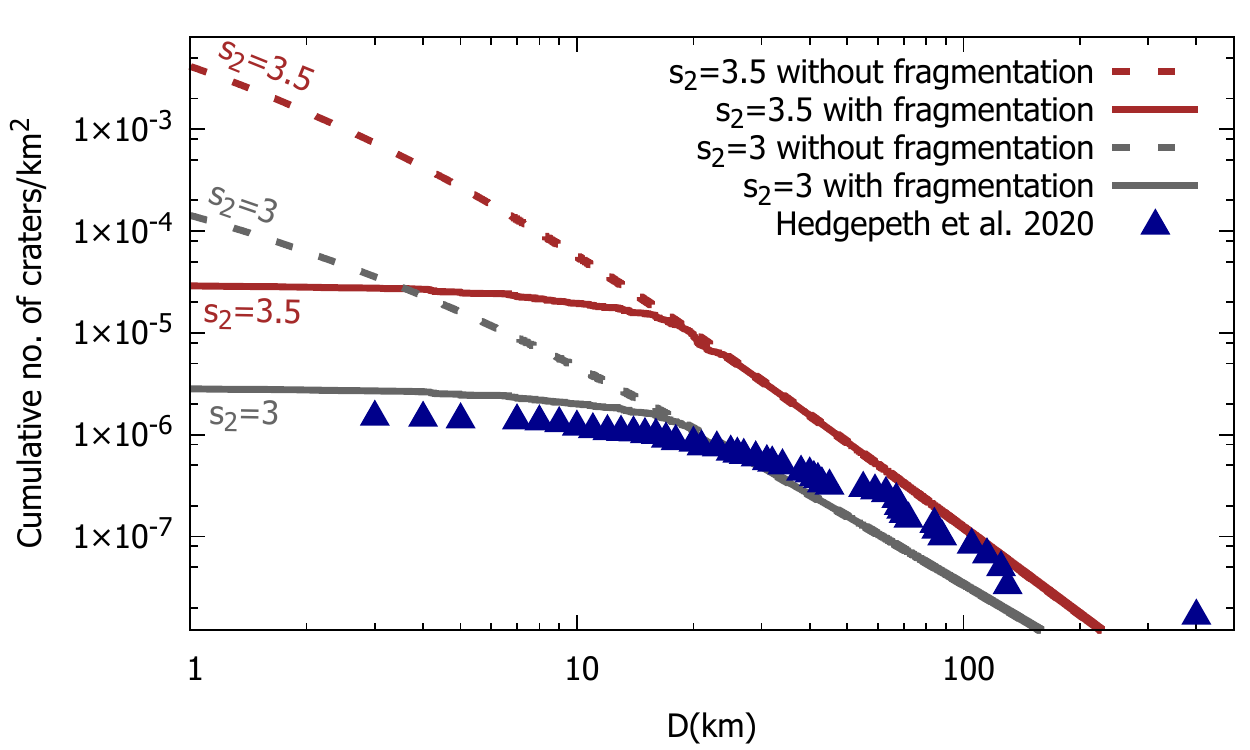}
\caption{Cumulative number of craters per square kilometer on Titan as a function of crater diameter, for two limiting values of the $s_2$ value. Solid lines correspond to the model that allows impactor fragmentation and dashed lines indicate the results when impactor fragmentation is not considered. The blue triangles represent crater counts from \citet{Hedgepeth2020}.}
\label{distroc}
\end{figure}

In Fig. \ref{distroc}, we present our predicted cumulative number of craters per square kilometer for both $s_2$ values, together with the observed crater counts from \citet{Hedgepeth2020}. 
Specific results for the case of the largest impactor for both $s_2$ indexes of the impactor size distribution are shown in Table \ref{t2}. 

\begin{table}[h!]
\caption{\label{t2}Model results.}
\begin{center}
{\begin{tabular}{|c|c|c|c|c|c|c|} 
\hline
\rule{0pt}{2ex} $D^{*}$ & $v_\text{f}$ & $d_m$ & $D_m$ & $v_\text{f}$ & $d_m$ & $D_m$ \\
& $s_2$=3 & $s_2$=3 & $s_2$=3 & $s_2$=3.5 & $s_2$=3.5 & $s_2$=3.5 \\
  \hline
\rule{0pt}{2.5ex} 2.11 & 7.29 & 12.88 & 158.12 & 7.31 & 19.4 & 227.01 \\
  \hline
\end{tabular}}
\tablefoot{Transition crater diameter $D^{*}$ between simple and complex craters (in kilometers); final vertical collision velocity $v_\text{f}$ of the largest impactor (in kilometers per second); largest impactor diameter $d_m$ and largest crater diameter $D_m$ on Titan for both $s_2$ indexes (in kilometers).}
\end{center}
\end{table}

As can be seen in Fig. \ref{distroc}, the results obtained with the $s_2=3.5$ index present the most similar fit to the observations for craters with diameters $D> 50$ km, with the exception of crater Menrva ($D=400$ km). A crater of this size may have formed early in the history of Titan, during the initial mass depletion that produced the late heavy bombardment of minor bodies on the planets, a process that is not included in our model. For craters with $D< 50$ km, our model with the $s_2=3.5$ index overestimates the number of craters even when a simple fragmentation model is considered. However, including impactor fragmentation and pancaking effects (solid lines in Fig. \ref{distroc}) produces a considerable change in our predicted crater distributions for those craters with $D \lesssim 25$ km, which flatten, emulating the same behavior as the observed distribution for that diameter range.

According to previous studies on Titan's atmospheric shielding effects, disruption becomes minimal at crater diameters of D $\geq 20$ km \citep{Neish2012, Artemieva2003, Korycansky2005}. Considering that only well-preserved small craters may be observable on Titan due to erosion and resurfacing \citep{Wood2010,Neish2012}, an overestimation of the number of small craters is to be expected. In particular, fluvial processes may be able to erode craters beyond recognition, explaining their scarcity in polar regions \citep{Neish2016}. Provided that several works consider the $s_2=3$ index for the size distribution of impactors with $d<100$ km \citep[e.g.,][]{Shoemaker1982}, we find that the crater distribution obtained via this index underestimates the number of craters for all craters with $D \gtrsim 25$ km (see Fig. \ref{distroc}). 

In addition, we  also studied the crater distribution resulting from the index $s_2=2.5$ considered in previous works \citep{DiSisto2011,DiSisto2013,Rossignoli2019}. However, for $s_2=2.5$ the results fall under the observed distribution for almost all crater diameters $D$, and thus have not been included in Fig. \ref{distroc}. 

\subsection{Comparison with other predicted crater distributions}
\begin{figure}
\includegraphics[width=1\linewidth]{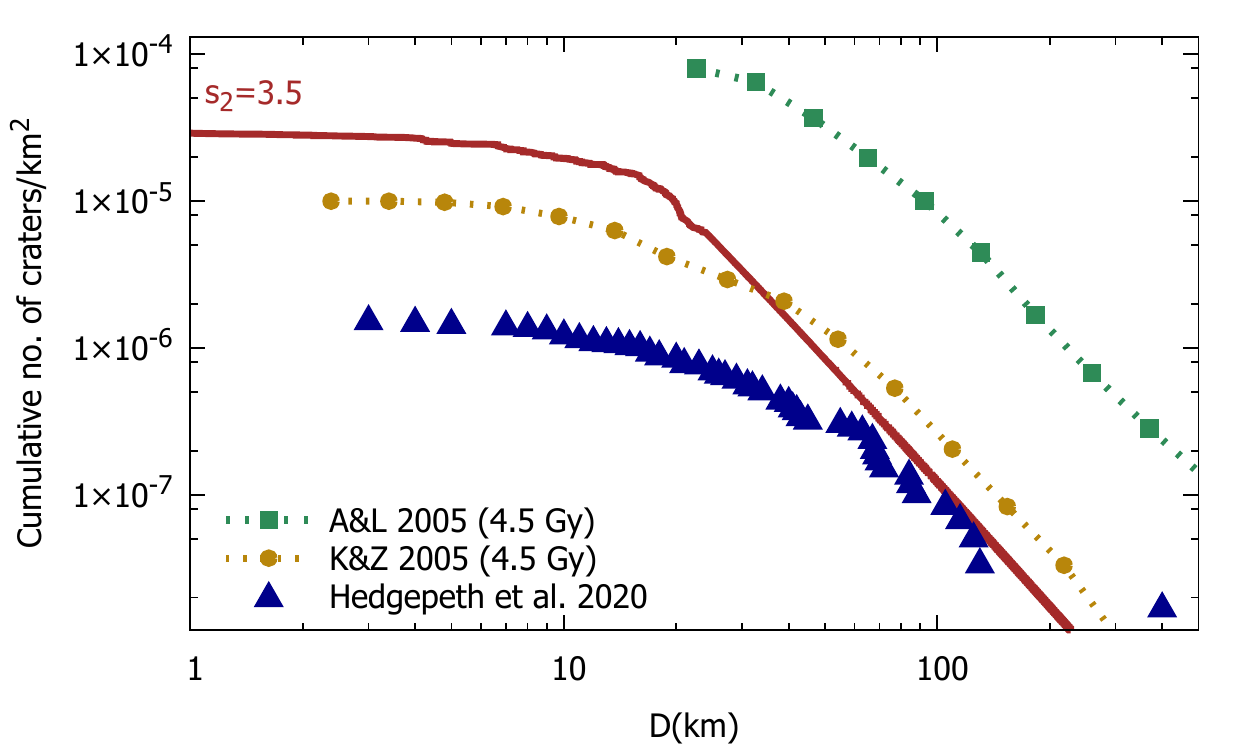}
\caption{Our model's predicted cumulative number of craters per square kilometer as a function of crater diameter for $s_2$ = 3.5, compared to results by AL05 and KZ05 for a 4.5 Gyr surface. The blue triangles correspond to crater counts from Hedgepeth (2020).}
\label{compa}
\end{figure}

\begin{figure}
\includegraphics[width=1\linewidth]{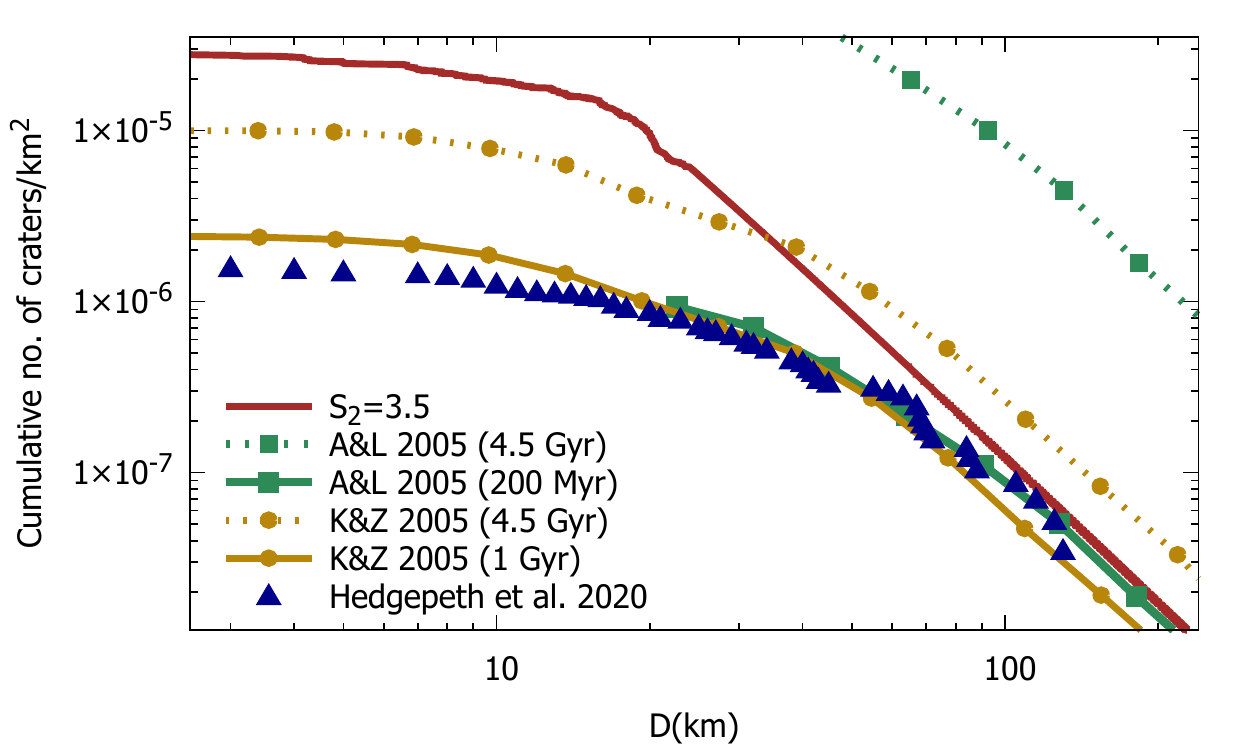}
\caption{Magnified image of Fig. 5 including results of AL05 and KZ05 for a 200 Myr and 1 Gyr surface, respectively.}
\label{compb}
\end{figure}

\begin{figure}
\includegraphics[width=1\linewidth]{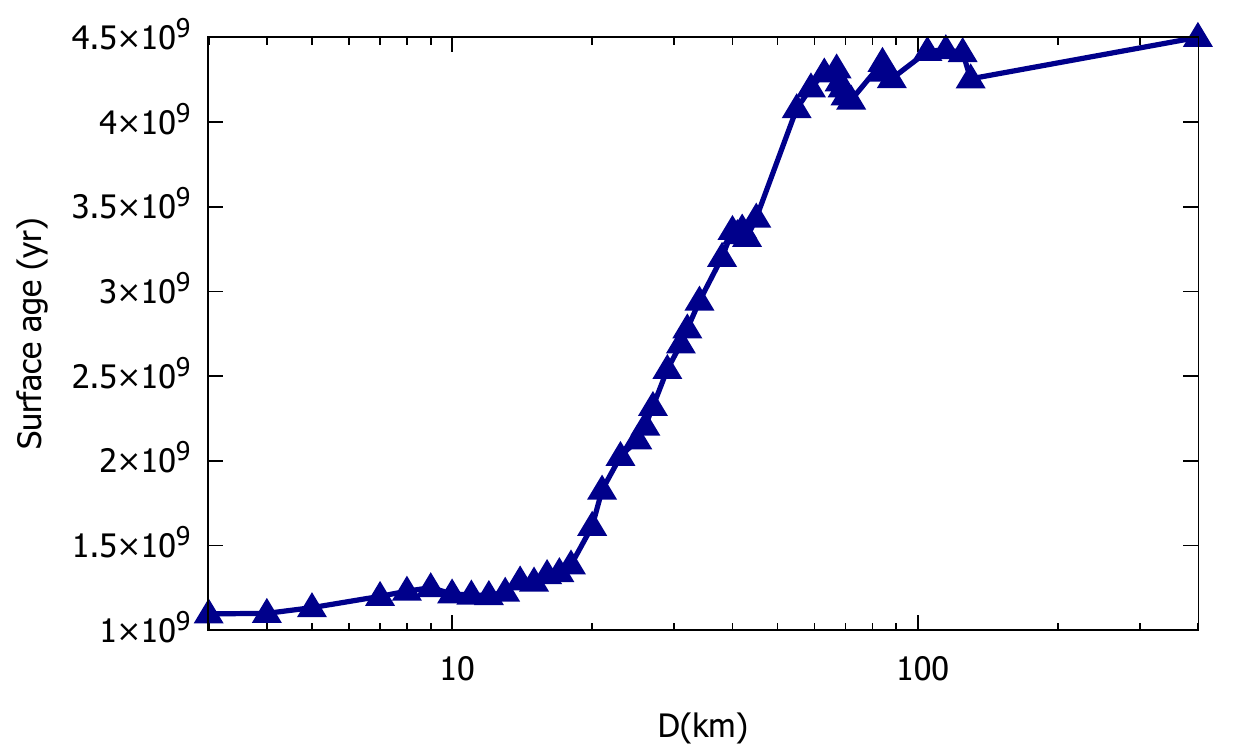}
\caption{Titan's surface age as a function of crater diameter $D$ for $s_2$ = 3.5. The blue triangles correspond to crater counts from Hedgepeth (2020).}
\label{age}
\end{figure}

In order to compare our results to other estimations of the crater size distribution on Titan we consider the models presented in \citet{Artemieva2005} (hereafter AL05) and KZ05. The results obtained in these papers were analyzed and compared thoroughly by \citet{Wood2010} and \citet{Neish2012}. The AL05 and KZ05 models both included impactor disruption by Titan's atmospheric shielding and used the impact rate presented in \citet{Zahnle2003}. However, KZ05 considered a constant impact rate, while AL05 assumed a $1/t$ dependence over time. The linear fit in Fig. \ref{encounters} allows us to obtain the current cratering rate on Titan and ease the comparison between our model and the cratering rates for Titan presented in \citet{Zahnle2003} and \citet{Dones2009}. The slope value $\dot{C}$ (see Sect. \ref{surf-age}) enables us to compute, for the past 3 Gyr, a cratering rate on Titan for craters with diameters $D>10$ km of $5.45\times10^{-15}$ yr$^{-1}$ for an airless Titan and of $1.28 \times10^{-15}$ yr$^{-1}$ if the atmospheric effects are included, all considering $s_2=3.5$ in Eq. (\ref{nr}). The corresponding value presented in Table 4 of \citet{Zahnle2003} for the case $A$ impactor population, where the size-number distribution of impactors is inferred from craters on the Galilean satellites, is $6\times10^{-15}$ craters per year with an uncertainty factor of four, while in \citet{Dones2009} the present-day cratering rate at an airless Titan for the case $A$ impactor population is  $3.4\times10^{-15}$ yr$^{-1}$. Another point of comparison between the KZ05 and AL05 models are their crater scaling laws, which was determined by \citet{Neish2012} to be the aspect where the most important differences between the models resided. On the one hand, AL05 considered the projectile density $\rho_{\text{i}}=1$ g cm$^{-3}$ (equivalent to our model), a scaling law for a water target, and ${\eta/(1-\eta)}=0.176$ for the scaling exponent in Eq. (\ref{dcrater}). On the other hand, in KZ05 the projectile density value was $\rho_{\text{i}}=0.5$ g cm$^{-3}$, the scaling law was for a sand target, and ${\eta/(1-\eta)=0.13}$. Therefore, these models led to two distinct predicted crater distributions on Titan, with a difference that ranges between a factor of two for small craters to a factor of 30 for craters of 1000 km \citep{Neish2012}. With respect to Titan's surface chronology \citet{Lorenz2007}, \citet{Wood2010}, and \citet{Neish2012} found that the results from KZ05 are most compatible with a crater retention age of 1 Gyr, while the crater distribution obtained in AL05 is more consistent with a crater retention age of 200 Myr \citep{Wood2010,Neish2012}. Figure \ref{compa} shows our results for the $s_2=3.5$ index with the model that includes impactor fragmentation, together with the results from AL05 and KZ05 for a 4.5 Gyr surface.
As can be seen, the crater distributions from AL05 and KZ05 considering craters formed over the entire Solar System age overestimate the number of craters. For this reason, \citet{Neish2012} presented the predicted crater distributions from AL05 and KZ05 adopting a crater retention age of 200 Myr and 1 Gyr, respectively (Fig. \ref{compb}). In contrast, our model with the $s_2=3.5$ index is able to predict consistently those craters with $D> 50$ km considering craters formed over the entire Solar System age and without requiring a complete global resurfacing between 200 Myr and 1 Gyr ago. 
Instead, our crater retention age calculation (Fig.~\ref{age}) shows that craters with $D>50$ km may be as old as the Solar System, which supports the idea that Titan could be a primordial object. On the other hand, craters with $D<10$ km reach ages of $\sim$ 1 Gyr. These results are similar to those presented in \citet{Lorenz1996}, which suggest that craters with $D=20$ km could be preserved for $\sim$ 2 Gyr, while craters with $D>50$ km are not likely to be eroded to the point of being undetectable.  

Crater retention ages presented in \citet{Wood2010} and \citet{Neish2012} are based on the best fit between the AL05 and KZ05 predicted crater distributions and the observed crater counts. This method leads to a global crater retention age that considers a normalized age for all crater sizes. Instead, our results for the surface age of different  Saturnian satellites \citep{DiSisto2016,Rossignoli2019} show that the crater retention age is dependent on the crater size. Larger craters tend to be preserved for a longer time and may be detectable even if affected by erosive processes, while smaller craters may be eroded beyond recognition on short timescales. On Titan smaller craters may be more easily obliterated by erosive processes such as fluvial modification \citep{Forsberg2004}, although atmospheric shielding effects and the uncertainty in crater counts for $D<20$ km prevent an accurate quantification of this type of erosion \citep{Neish2016}. Nevertheless, our results for   Titan's 
 crater retention age show the same size dependence as was found for other Saturnian satellites \citep{DiSisto2016,Rossignoli2019}. Thus, our size-dependent chronology results may help to constrain different erosion processes acting on Titan's surface.     

\subsection{Apex-antapex asymmetry}
The hemispheric distribution of craters produced on a synchronously rotating satellite has been studied in several works \citep{Shoemaker1982,Horedt1984,Zahnle2001} and is predicted to be denser on the leading side of the satellite for the case of heliocentric impactors. However, for most of the satellites in the outer Solar System no distinct asymmetries have been found  \citep[e.g.,][]{Zahnle2001,Dones2009,Kirchoff2010}.  \citet{Wood2010} analyzed the hemispheric distribution of the 49 impact craters on Titan detected at the time and found that 63\% of the total number of craters occurred on the leading side. This exact percentage is still valid when analyzing the updated set of 90 observed craters listed in \citet{Hedgepeth2020}. Following the method presented in \citet{Shoemaker1982} and using the velocities listed in Table \ref{tabladatos}, we find that the expected leading/trailing asymmetry depends strongly on the hemispheric difference in the modeled mean impact velocity at the top of the atmosphere. Our results using \citet{Shoemaker1982} derivations give mean impact velocities at the leading and trailing sides of 13.59 km\,s$^{-1}$ and 7.17 km\,s$^{-1}$, respectively, which results in a cratering rate asymmetry of 15:1. In contrast, if we use the leading and trailing sides mean velocities modeled in KZ05, which are 10.9 km\,s$^{-1}$ and 9.4 km\,s$^{-1}$, respectively, 
we obtain a cratering rate asymmetry of 4.9:1, very similar to the 4:1 ratio obtained in KZ05 and the 5:1 ratio from \citet{Lorenz1997}. Nevertheless, even the lowest predicted cratering rate asymmetry is higher than the observed 1.7:1 ratio. This difference may be due to a number of factors. On the one hand, there is the possibility that Titan has not rotated synchronously for the age of the Solar System. On the other hand, a planetocentric population of impactors such as fragments from the putative breakup of the Hyperion parent body \citep{Farinella1990} may have altered the expected hemispheric cratering asymmetry, as Titan may be able to accrete $\sim$ 78\% of Hyperion’s ejecta \citep{Dobrovolskis2004}.

\section{Conclusions}
\label{SecConclusions}   
In the present work we  modeled the impact crater distribution on Titan generated by centaurs over the Solar System age. In order to obtain the number of impacts on Titan we  used the results from an updated simulation of the dynamical evolution of SDOs and their contribution to the centaur population \citep{DiSisto2020}. The CSD of SDOs was modeled considering the current uncertainties in the size distribution for the smaller objects of the impactor population. Thus, we   considered two limiting values for the differential power law index for objects with $d$ < 100 km, $s_2=3$ and $s_2=3.5$, and our results for the impact crater distribution are presented for both of these values. In addition, we  incorporated a simple model for the atmospheric shielding of impactors where the effects of fragmentation, pancaking, deceleration, and ablation are included considering the current density profile for Titan's atmosphere throughout the entire simulation. Last, we  compared our results with the most updated crater counts from {\em Cassini} \citep{Hedgepeth2020} and with the predicted crater distributions presented in \citet{Neish2012} by AL05 and KZ05.  

Our results show that the predicted crater distribution obtained with the $s_2=3.5$ index is more consistent with the observed crater distribution, especially for larger craters ($D>50$ km) which are less affected by erosion. For craters with $D<50$ km, our model overestimates the number of craters and the difference between the theoretical and observed distributions grows larger for smaller sizes down to $D \sim 25$ km, where both the observed and predicted crater distributions become flatter. This similar behavior may indicate that our model is able to correctly constrain the atmospheric shielding effects. Thus, the difference between our predicted cratering rates with the $s_2=3.5$ index and the observations can be considered a measure of the extent to which erosive processes have acted on Titan's surface throughout the Solar System age. In this respect, our calculations on the crater retention age show that Titan's surface is able to retain evidence of its largest craters
  over the
  age of the Solar System, while the smallest craters may be eroded beyond detection on timescales of $\sim$ 1 Gyr.

\begin{acknowledgements}
NLR and RPDS are grateful for support from IALP and Agencia de Promoción Científica y Tecnológica through the PICT 201-0505. MGP thanks CONICET for partial support through the research grant PIP 112-201501-00699CO. This work was partially supported by Universidad Nacional de La Plata (UNLP) through PID G172. The authors wish to thank Catherine Neish and Alessandro Morbidelli for valuable discussions and comments and Emma Fernández Alvar for providing an essential piece of bibliography for this work. We thank R. Lorenz and an anonymous referee for their helpful comments and corrections which helped us to improve this work.  
\end{acknowledgements}

%
%
\bibliographystyle{aa}
\bibliography{biblio}
\end{document}